\documentclass[preprint,12pt]{elsarticle}
\usepackage[dvipsnames]{xcolor}
\usepackage[T1]{fontenc}
\usepackage[utf8]{inputenc}
\usepackage{caption}
\usepackage[colorlinks = true]{hyperref}
\usepackage{amsmath,amsfonts,amssymb}
\usepackage{bm}
\usepackage{url}
\makeatletter
\g@addto@macro{\UrlBreaks}{\UrlOrds}
\makeatother
\usepackage{mmacells}

\usepackage{slashed}
\usepackage{xspace}
\usepackage{longtable}

\journal{Computer Physics Communications}

\newcommand{\mma}{\textsc{Mathematica}\xspace}
\newcommand*{\cf}{cf.\ }
\newcommand*{\eg}{e.\,g.\ }
\newcommand*{\ie}{i.\,e.\ }

\mmaSet{morefv={gobble=2},morelst={breaklines=true}}

\begin{document}

\begin{frontmatter}

\title{FeynCalc 10: Do multiloop integrals dream of computer codes?}

\author[a,b]{Vladyslav Shtabovenko\corref{author1}}
\author[c]{Rolf Mertig\corref{author2}}
\author[d]{Frederik Orellana\corref{author3}}

\cortext[author1] {\textit{E-mail address:} shtabovenko@physik.uni-siegen.de}
\cortext[author2] {\textit{E-mail address:} rolfm@gluonvision.com}
\cortext[author3] {\textit{E-mail address:} fror@dtu.dk}

\address[a]{Theoretische Physik 1, Center for Particle Physics Siegen,  
	Universität Siegen, \\ Walter-Flex-Str. 3, 57068 Siegen, Germany}
\address[b]{Institut für Theoretische Teilchenphysik (TTP),
	Karlsruhe Institute of Technology (KIT), 
	\\
	Wolfgang-Gaede-Straße 1, 76131 Karlsruhe, Germany}
\address[c]{GluonVision GmbH, B\"otzowstr. 10, 10407 Berlin, Germany}
\address[d]{Technical University of Denmark, Anker Engelundsvej 1, \\
	2800 Kgs. Lyngby, Denmark}

\begin{abstract}
In this work we report on a new version of \textsc{FeynCalc}, a  \mma package widely 
used in the particle physics community for manipulating quantum field theoretical expressions and
calculating Feynman diagrams. Highlights of the new version include greatly improved capabilities for doing
multiloop calculations, including topology identification and minimization, optimized tensor reduction, rewriting of scalar products in terms of inverse denominators, detection of equivalent or scaleless loop integrals, derivation of Symanzik polynomials, Feynman parametric as well as graph representation for master integrals and initial support for handling differential equations and iterated integrals. In addition to that, the new release also features completely rewritten routines for color algebra simplifications, inclusion of symmetry relations between arguments of Passarino--Veltman functions, tools for determining matching coefficients and quantifying the agreement between numerical results, improved export to \LaTeX \, and first steps towards a better support of calculations involving light-cone vectors.
\end{abstract}

\end{frontmatter}

{\bf PROGRAM SUMMARY/NEW VERSION PROGRAM SUMMARY}

\begin{small}
\noindent
{\em Program Title:} FeynCalc                                         \\
{\em CPC Library link to program files:} (to be added by Technical Editor) \\
{\em Developer's repository link:} \url{https://github.com/FeynCalc/feyncalc} \\
{\em Code Ocean capsule:} (to be added by Technical Editor)\\
{\em Licensing provisions:} GPLv3 \\
{\em Programming language:} Wolfram Language                                  \\
{\em Supplementary material:} Manual, example notebooks                        \\
{\em Journal reference of previous version:} Comput. Phys. Commun. \textbf{256} (2020) 107478\\
{\em Does the new version supersede the previous version?:} Yes  \\
{\em Reasons for the new version:} Addition of new routines required for multiloop calculations. \\
{\em Summary of revisions:} \textsc{FeynCalc} can be now used to calculate multiloop Feynman diagrams either standalone or as a part of a toolchain.\\
{\em Nature of problem:} Analytic calculations of higher-order corrections to particle physics processes using Feynman diagrammatic expansion.\\
{\em Solution method:} The required algorithms and algebraic identities are implemented in Wolfram \mma.\\
{\em Restrictions:} Depending on the complexity of the problem, the number of terms might become so high that \mma alone will not be sufficient to finish the calculation within a reasonable time frame or at all.

\end{small}

\section{Introduction}

Modern high energy physics heavily relies on Feynman's diagrammatic approach to the calculation of perturbative corrections in particle reactions. The vast majority of predictions required to match the expected experimental precision of the High luminosity LHC \cite{Apollinari:2015wtw} as well as the proposed future colliders are still obtained by calculating multitudes of complicated Feynman diagrams, and it does not seem likely, that this situation is going to change in the near future.

The number of diagrams involved can easily go into thousands or even millions, which makes the usage of automation indispensable. The importance of efficient algorithms for such calculations has already been recognized decades ago (\cf \eg refs.~\cite{Veltman:1972rt,Tkachov:1996wh}) and the unceasing development of new calculational techniques is one of the main reasons why calculations unthinkable now might nonetheless become feasible in the next years. A good overview over modern methods used in higher-order perturbative calculations can be found \eg in refs.~\cite{Smirnov:2006ry,Weinzierl:2022eaz}.

While efficient codes for automatic numerical calculations at tree-level and one-loop accuracy are available to the wide public since years, the treatment of multiloop diagrams remains challenging and requires considerable expertise.

Beyond one-loop the proliferation of diagrams becomes an issue, where processes involving thousands or tens of thousands diagrams are not uncommon. Their complexity increases as well, meaning that algebraic evaluation requires significantly more time and computational resources. At one loop all integrals with quadratic propagators can be conveniently reduced to a basis (\eg that of Passarino--Veltman~\cite{Passarino:1978jh}) and then evaluated either analytically or numerically using existing libraries. However, already at two loops this approach becomes unfeasible.
First, the complete basis of two-loop master integrals with arbitrary masses and multiple legs is not known yet.
Second, the analytic reduction of multiloop integrals to some basis integrals is currently doable only when they involve few kinematic scales. For example, as of now a reduction of integrals appearing in the $2 \to 3$ process $q \bar{q} \to t \bar{t} H$ with five scales requires an intricate combination of state-of-the-art numerical and analytic techniques \cite{Agarwal:2024jyq} and cannot be done by brute force. At the same time, at one-loop a diagram with five legs and five scales would not pose any major difficulties. Even when one manages to obtain a final set of master integrals, their evaluation often turns out to be a significant challenge. Analytic results are scarce \cite{Bogner:2017xhp} and are often not generic enough to cover all phenomenologically interesting (\ie those with different masses and multiple legs) master integrals even at two loops. Numerical evaluation using sector decomposition \cite{Hepp:1966eg,Speer:1977uf,Binoth:2000ps,Heinrich:2008si} is always possible\footnote{For some problematic integrals one may need to request help from the code developers, though.}, but achieving sufficient precision in problematic phase space regions can quickly become challenging. In this sense, performance issues arising in multiloop calculations are usually solved on a case-by-case basis, but for a general-purpose code one would need better algorithms that are efficient enough to cover all interesting processes, in particular those involving massive particles.

Limiting ourselves to multiloop calculations that are feasible with the current-day technology, the task of streamlining all the necessary calculational steps such, that they can be performed in a fully automatic fashion, is still very much work in progress. A recent review of the field can be found \eg in ref.~\cite{Campbell:2022qmc}. Some remarkable achievements towards multiloop automation are the release of \textsc{Caravel}~\cite{Abreu:2020xvt}, a framework for multiloop computations using numerical unitarity, as well as the recent version of \textsc{pySecDec}~\cite{Heinrich:2023til} capable of evaluating multiloop amplitudes numerically. Apart from that, there are also ongoing efforts to ``upgrade'' existing one-loop codes at least to two-loop accuracy (\cf \eg \cite{Borowka:2016agc,Borowka:2016ehy,Pozzorini:2022ohr,Zoller:2022ewt,Canko:2023lvh}).

Irrespective of whether one wants to calculate a cross section, a matching coefficient or a renormalization constant, the vast majority of multiloop calculations usually require the completion of some fundamental steps that can be summarized as follows: (i) generation of Feynman diagrams for the given process, (ii) algebraic simplification of the corresponding amplitudes (including suitable expansions in small parameters), (iii) reduction of the occurring loop integrals to a smaller set of master integrals, using Integration-By-Parts (IBP)~\cite{Chetyrkin:1981qh,Tkachov:1981wb} techniques, (iv) evaluation of the master integrals. Fortunately, the first three steps in the above list can be completed in a highly automatized fashion by combining publicly available software with suitable self-written code. In most cases the problems one has to deal with happen to be of technical (\eg bugs in the code, performance bottlenecks, lack of computing power) rather than conceptual nature. Analytic calculations of master integrals usually require more experience and creativity, unless the desired results are already available in the literature and can be directly plugged into the final result. 

A lot of frameworks addressing steps (i)-(iii) happen to be some private codes developed by single researchers or whole research groups specializing on loop calculations. Sometimes they can be  obtained upon request (albeit without much support or proper documentation) but most tools are still available only to collaborators. With the new generation of researchers embracing open source ideas and making their codes public (\cf \eg refs.~\cite{Gerlach:2022qnc,Maheria:2022dsq,Feng:2021kha,Wu:2023qbr}) this situation started to change. The corresponding tools aim at connecting different steps behind a multiloop calculation to each other within a single framework and normally need to address the tasks of generating Feynman diagrams, inserting Feynman rules, identifying the occurring loop integral topologies, minimizing their number and supplying some templates for the subsequent amplitude evaluation in \textsc{FORM}~\cite{Vermaseren:2000nd,Kuipers:2012rf}.

One aim of the present work is to  make analytic multiloop calculations accessible to a broader range of particle theorists. The method is to extend the well-established one-loop functionality of \textsc{FeynCalc}~\cite{Mertig:1990an,Shtabovenko:2016sxi,Shtabovenko:2020gxv,Shtabovenko:2021hjx} with the modern state-of-the art algorithms. The new version of the package presented here thus includes a large set of optimized routines for dealing with multiloop calculations.

This paper is the first in a series of three publications describing our take on creating a new framework for semi-automatic multiloop calculations. While this work revolves solely around \textsc{FeynCalc}, the two subsequent papers will introduce a new version of \textsc{FeynHelpers} ~\cite{Shtabovenko:2016whf} --- an add-on for connecting \textsc{FeynCalc} to other software tools related to quantum field theory (QFT) and finally a \textsc{FORM}-based framework for symbolic evaluation of Feynman diagrams that makes use of \textsc{FeynCalc} and \mma during certain calculational steps.

The remainder of the present paper is organized as follows: In Section \ref{sec:vision} we briefly introduce \textsc{FeynCalc} and compare it to similar codes, while instructions on installing the package from \textsc{GitHub} can be found in Section \ref{sec:install}. New symbols and functions related to the multiloop capabilities of the package are presented in Section \ref{sec:topos}. Section \ref{sec:masters} discusses some special routines that can be handy when calculating master integrals. Last but not least, new features and improvements that are not directly related to multiloop calculations can be found in Section \ref{sec:rest}. A description of example calculations utilizing new multiloop capabilities is offered in Section \ref{sec:examples}. Finally, in Section \ref{sec:summary} we summarize our results and provide an outlook on the future of the package.

Contrary to previous \textsc{FeynCalc} papers, the amount of \mma code presented will be kept to a bare minimum. The reason is that \textsc{FeynCalc} 10 features a comprehensive manual, covering every symbol introduced in the package (including examples) and containing a tutorial for learning the basics of the package. In addition to that, examples related to the functionality described in this paper can be found in the accompanying \mma notebook.

\section{Context and state of the art} \label{sec:vision}

Originally, \textsc{FeynCalc} was designed to handle loop integrals using Passa\-rino-Veltman \cite{Passarino:1978jh} reduction, which effectively limited its applicability to one-loop calculations only. Subsequent iterations of the code \cite{Mertig:1995ny} allowed for working with certain types of multiloop integrals (\eg using the \textsc{TAR\-CER}~\cite{Mertig:1998vk} add-on) but until version 10 it was neither very efficient nor suitable for general purpose multiloop calculations.

The number of actively developed tools for semi-automatic calculations with an emphasis on tree- and one-loop level similar to \textsc{FeynCalc} has been steadily decreasing over the last years. Both \textsc{HepMath} \cite{Wiebusch:2014qba} and \textsc{Package-X}~\cite{Patel:2015tea,Patel:2016fam} have unfortunately been abandoned, while there seems to be hardly new codes covering a wide range of applications as well as receiving regular updates and bug fixes. In this context it is also worth mentioning \textsc{Physics}\footnote{\url{https://www.maplesoft.com/products/maple/features/physicsresearch.aspx}}, a package shipped with \textsc{Maple} computer algebra system, whose particle physics related capabilities have been significantly extended in the last few years. Also, \textsc{formtracer} \cite{Cyrol:2016zqb} can be handy in some circumstances, although this tool is mostly limited to algebraic simplifications.

What we regard as a crucial feature of semi-automatic codes is the ability to access intermediate expressions at any stage of the calculation and to organize calculations in a flexible way by sequentially applying high-level functions to the original input. Depending on the quantities one wants to derive, this approach might be more efficient than using programs offering a higher level of automation and less flexibility. Of course, fully automated tools for tree-level and one-loop calculations such as \textsc{MadGraph}~\cite{Alwall:2014hca}, \textsc{GoSam}~\cite{Cullen:2011ac,GoSam:2014iqq}, \textsc{Herwig}~\cite{Bahr:2008pv,Bellm:2015jjp}, \textsc{HELAC-NLO}~\cite{Bevilacqua:2011xh}, \textsc{POWHEG-BOX}~\cite{Nason:2004rx,Frixione:2007vw,Alioli:2010xd}, \textsc{Sherpa}~\cite{Gleisberg:2008ta,Sherpa:2019gpd}, \textsc{Whizard} \cite{Moretti:2001zz,Kilian:2007gr}, \textsc{CalcHep}~\cite{Belyaev:2012qa}, \textsc{CompHep}~\cite{CompHEP:2004qpa}, \textsc{GRACE} \cite{Yuasa:1999rg,Fujimoto:2002sj} etc. are very useful when performing the tasks they were originally designed for, \eg calculating cross-sections and decay rates.

Since version 10 \textsc{FeynCalc} can be in principle used to perform real multiloop calculations or at least to derive multiloop amplitudes written as linear combinations of loop integrals belonging to a certain set of integral topologies. Of course, due to the performance limitations of \mma as compared to \textsc{FORM} the number of multiloop diagrams one can completely evaluate with \textsc{FeynCalc} and their complexity are rather limited. Thus, in this context, \textsc{FeynCalc} should be seen as providing supplementary tools for structuring and reducing multiloop calculations and a supplementary way to check specific calculations of other multiloop frameworks. Our code has already been employed in this fashion in some research projects \cf \eg \cite{Gerlach:2022hoj, 
Stockinger:2023ndm,Yang:2023acc,Gehrmann:2023ksf,Kuhler:2024fak,Reeck:2024hdn,Akpinar:2024meg} and we expect more publications to appear in the near future.

\textsc{Alibrary}\footnote{\url{https://magv.github.io/alibrary/}} is one of the few tools that uses \mma for gluing different parts of the computational setup together and implementing some convenience functions. Nevertheless, this code uses \textsc{QGRAF}~\cite{Nogueira:1991ex} for generating Feynman diagrams (Feynman rules for QCD are already included)  and \textsc{FORM} for evaluating them. The topology identification part is outsourced to a dedicated tool called \textsc{feynson}~\cite{Maheria:2022dsq} --- developed by the author of \textsc{Alibrary}. The package also includes interfaces to \textsc{GraphViz}~\cite{url:Graphviz}, \textsc{LiteRed}~\cite{Lee:2012cn,Lee:2013mka}, \textsc{FIRE}~\cite{Smirnov:2014hma,Smirnov:2019qkx,Smirnov:2023yhb}, \textsc{KIRA}~\cite{Maierhofer:2017gsa,Maierhofer:2018gpa,Maierhofer:2019goc,Klappert:2020nbg,Lange:2021edb}, \textsc{pySecDec}~\cite{Borowka:2017idc,Borowka:2018goh,Heinrich:2021dbf} and other related programs. \textsc{FORM} source files or configuration files for \textsc{FIRE} and \textsc{KIRA} generated by \textsc{Alibrary} can be used independently --- which is an approach similar to that of \textsc{FeynCalc}. Tensor reduction and Dirac algebra simplifications are not included, since the code tacitly assumes the usage of projectors. However, those parts can be still added to the generated \textsc{FORM} code by hand at a later stage.

The original idea behind the  \textsc{Python} package \textsc{tapir} \cite{Gerlach:2022qnc} was to create a modern replacement for \textsc{q2e}~\cite{Harlander:1998cmq,Seidensticker:1999bb}, a \textsc{C++} code for inserting Feynman rules into \textsc{QGRAF} output. Together with \textsc{exp} and \textsc{calc}, \textsc{q2e} is a part of the so-called Karlsruhe tool chain\footnote{The tools \textsc{q2e}, \textsc{exp} and \textsc{calc} are not public, but can be obtained upon request, \cf \url{http://sfb-tr9.ttp.kit.edu/software/html/q2eexp.html}} that has been used in many cutting-edge multiloop calculations. In the course of its development, \textsc{tapir} obtained numerous features that go far beyond the capabilities of \textsc{q2e}. In particular, it can be also used for identifying and minimizing integral topologies, visualizing Feynman diagrams, performing partial fraction decomposition and generating amplitudes in \textsc{FORM} format. Furthermore, it understands Feynman rules in the Universal FeynRules Output (UFO)~\cite{Degrande:2011ua,Darme:2023jdn} format. The 
\textsc{FORM} code for evaluating the amplitudes is, however, not part of \textsc{tapir}. Just as in the case of \textsc{Alibrary}, \textsc{tapir} has no built-in capabilities to perform tensor reduction of loop integrals or to simplify chains of Dirac matrices that do not involve Dirac traces.

The program \textsc{FeAmGen.jl}~\cite{Wu:2023qbr} is written in the \textsc{Julia} language, has built-in support for UFO models and uses \textsc{YAML} run cards describing the process that needs to be calculated. Automatically generated \textsc{FORM} code takes care of Dirac and color algebra, while a built-in routine of \textsc{FeAmGen.jl} minimizes the number of loop integral topologies. However, the evaluated amplitude still requires tensor reduction or a suitable projector, while the obtained topologies are not readily converted into configuration files for IBP-reduction tools such as \textsc{FIRE} or \textsc{KIRA}. On the other hand, since the output of \textsc{FeAmGen.jl} is offered in form of \textsc{JDL2} or plain text files, those additional steps can be also done by the user.

In the case of \textsc{HepLib}~\cite{Feng:2021kha,Feng:2023hxy}, the authors chose to employ \textsc{C++} and in particular the \textsc{GiNaC} \cite{Bauer:2000cp} library as the means of connecting different calculational steps with each other. As such, this tool is usable both in \textsc{C++} and \textsc{Python}. Apart from the fact that \textsc{HepLib} uses \textsc{QGRAF} and \textsc{FORM} for the common tasks of generating and evaluating amplitudes, it also features tensor reduction, partial fraction decomposition and automatic creation of configuration files for \textsc{FIRE} and \textsc{KIRA} as well as a custom implementation of sector decomposition \cite{Hepp:1966eg,Speer:1977uf,Binoth:2000ps,Heinrich:2008si} for numerical evaluation of master integrals.  

{
	\scriptsize
	\renewcommand{\arraystretch}{1.3}
	\begin{longtable}[b]{|p {1.5cm} |p {2.0cm}|p {1.6cm}|p {1.6cm}|p {1.6cm}|p {1.6cm}|}		
		\hline
		Feature& {\scriptsize \textsc{FeynCalc}}   & \textsc{Alibrary} & \textsc{TAPIR} & \textsc{FeAmGen.jl} & \textsc{HepLib} \\
		\hline
		Language & \textsc{Mathe\-matica} & \textsc{Mathe\-matica} & \textsc{Python} & \textsc{Julia} & \textsc{C++} \\
		\hline
		Diagram generation & \textsc{FeynArts}, \textsc{QGRAF} (\textsc{FeynHelpers}) & \textsc{QGRAF} & \textsc{QGRAF} & \textsc{QGRAF} & \textsc{QGRAF} \\
		\hline
		Diagram visualization & \textsc{FeynArts}, \textsc{TikZ-Feynman} (\textsc{FeynHelpers}) & \textsc{TikZ}, \textsc{GraphViz} & \textsc{TikZ-Feynman} & \textsc{TikZ-Feynman} & \textsc{TikZ-Feynman} \\
		\hline
		Topology mappings & yes & \textsc{feynson} & yes & yes  & yes \\
		\hline
		Partial fractioning & yes & yes & yes & no & yes \\
		\hline
		Dirac algebra except traces & yes & no & no & yes & yes \\
		\hline
		Color algebra & yes & \textsc{color.h} & \textsc{color.h} & yes & yes \\
		\hline
		Tensor \phantom{xx} reduction & yes & no & no & no & yes \\
		\hline		
		Uses \textsc{FORM} & no & yes & yes & yes & yes \\
		\hline
		Interface to IBP \phantom{xx} codes & \textsc{FeynHelpers} & yes & no & yes & yes \\	
		\hline
		New  \phantom{xx} models & \textsc{FeynRules} & by hand  & \textsc{UFO} & \textsc{UFO} & by hand \\
		\hline
		\caption{Some differences between tools for automatizing multiloop amplitude evaluation.}
		\label{tab:tools}
	\end{longtable}
}

A comparison of some loop-related features present in \textsc{FeynCalc} and other tools is presented in Table~\ref{tab:tools}. It goes without saying that since all the above-mentioned codes employ \textsc{FORM}, they easily outperform \mma (and hence \textsc{FeynCalc}) when it comes to the evaluation of Feynman diagrams. \textsc{FeynCalc} can identify the occurring topologies and minimize their number as well as directly simplify Dirac and color algebra and carry out tensor reduction of loop integrals. Feynman diagram generation is done using \textsc{FeynArts}, although an experimental interface to \textsc{QGRAF} is available in the development version of the yet unreleased \textsc{FeynHelpers} add-on. The same goes for automatic generation of run cards needed to perform an IBP-reduction of the master integrals: This feature is not part \textsc{FeynCalc} but will be offered in near future via \textsc{FeynHelpers}.

\section{Installation} \label{sec:install}

The fastest and most convenient way to install \textsc{feyncalc} is to use the automatic installer by evaluating
{\small
	\begin{mmaCell}[index=1,moredefined={InstallFeynCalc}]{Input}
		Import@"https://raw.githubusercontent.com/FeynCalc/feyncalc/master/install.m"
		InstallFeynCalc[]
	\end{mmaCell}
}
\noindent on a freshly started \mma kernel. All versions of \mma from 10.0 upwards are supported. The code above will install the stable version of the package. The development version --- with potential bugs but also the newest features, can be
obtained by setting the option \texttt{InstallFeynCalcDevelopmentVersion} of \texttt{InstallFeynCalc} to \texttt{True}. In the case of internet connection problems one can also install the package manually. For further instructions we refer to the section ``Installation'' of the package manual. The source code of \textsc{FeynCalc} can be obtained from \url{https://github.com/FeynCalc/feyncalc}.

\section{Topologies and loop integrals} \label{sec:topos}

At one loop, almost every calculation involving only integrals with quad\-ratic propagators can be handled using the so-called Passarino--Veltman (PaVe) technique. By considering the most generic basis of tensor structures for the given rank made of metric tensors and external momenta, each occurring tensor integral can be reduced to a linear combination of scalar functions. These quantities are known as PaVe coefficient or scalar functions and can be straightforwardly evaluated analytically or numerically. Rewriting the amplitude in terms of these functions usually concludes the loop-related part of the calculation. PaVe-Reduction is implemented in numerous codes including \textsc{FeynCalc} --- where the relevant routines are called \texttt{TID} and \texttt{PaVeReduce}. The conceptual simplicity behind the PaVe technology and the availability of reliable numerical codes (\cf \eg refs.~\cite{Hahn:1998yk,Denner:2016kdg,Denner:2002ii,Denner:2005nn,Denner:2010tr,Ellis:2007qk,vanHameren:2010cp}) make it the default choice for the majority of practitioners. Unfortunately, the complexity of multiscale multiloop integrals does not allow one to apply these methods beyond one-loop with the same ease and efficiency. Instead, it is customary to treat each integral family on its own, by first reducing all relevant integrals to a smaller set of master integrals and then calculating those using suitable analytic or numerical techniques.

To that aim it is necessary to have code(s) that can (i) introduce integral families from analyzing the propagators present in the amplitude, (ii) minimize the number of integral families by finding possible mappings between them, (iii) ensure that the set of propagators in each family forms a basis and, if necessary, (iv) perform tensor reduction or (v) partial fraction decomposition. Upon completing these steps, one should obtain a list of integral topologies present in the amplitude and the corresponding loop integrals belonging to these topologies. This information can be then passed to an IBP-reduction program such as \textsc{FIRE}, \textsc{KIRA}, \textsc{LiteRed}, \textsc{Reduze}~\cite{Studerus:2009ye,vonManteuffel:2012np}, \textsc{AZURITE}~\cite{Georgoudis:2016wff} etc. that will minimize the number of integrals that need to be evaluated. In the following we would like to focus on explaining how these five steps can be performed with the aid of \textsc{FeynCalc}.

\subsection{Three main building blocks}

The three main building blocks of \textsc{FeynCalc}'s new multiloop capabilities are the symbols \texttt{FCTopology} and \texttt{GLI} as well as the routine \texttt{FC\-Feynman\-Prepare}. In this context \texttt{FCTopology} represents an integral family that consists of propagators forming a basis. The syntax reads
{\small
	\begin{mmaCell}[indexed=false,moredefined={FCTopology}]{Code}
		FCTopology[id, {propagators}, {loop momenta}, {external momenta}, {kinematics}, {extra}]
	\end{mmaCell}
}
\noindent where the first argument denotes the name of the topology, the second argument enumerates the propagators, while the third and fourth lists contain names of loop and external momenta. \texttt{id} can be a string or a symbol, while \texttt{\{propagators\}} must be a list of \texttt{FAD}, \texttt{SFAD}, or \texttt{GFAD} propagator containers.

Kinematic constraints (\eg specific values of scalar products made of external momenta) can be specified in the fifth argument, while the last argument may be used to incorporate some addition information (\eg that this topology is a subtopology of a larger topology). A simple \texttt{FCTopology} example would be
{\small
	\begin{mmaCell}[indexed=false,moredefined={FCTopology,SFAD,SPD}]{Code}
		FCTopology[topo1, {SFAD[p1], SFAD[p2], SFAD[q - p1 - p2], SFAD[q - p2], SFAD[q - p1]}, {p1, p2}, {q}, {Hold[SPD][q] -> qq}, {}]
	\end{mmaCell}
}
\noindent 
that describes an on-shell 2-loop propagator-type topology. Notice that the construction \texttt{Hold[SPD][q]} prevents the scalar product $q^2$ from being immediately evaluated in case it has already been assigned a value via \texttt{SPD[q] = xyz}.

Having defined an integral family, we can also introduce integrals belonging to it. To that aim \textsc{FeynCalc} uses \texttt{GLI}s (a shortcut for ``Generic Loop Integral'') defined as
{\small
	\begin{mmaCell}[indexed=false,moredefined={GLI}]{Code}
		GLI[id, {powers}]
	\end{mmaCell}
}
\noindent 
where the first argument denotes the family name (and must match the \texttt{id} of the corresponding \texttt{FCTopology}) and the second contains powers of the involved propagators. The propagator powers must integers, e.g.
{\small
	\begin{mmaCell}[indexed=false,moredefined={GLI}]{Code}
		GLI[topo1, {1,1,1,-2,2}]
	\end{mmaCell}
}
\noindent 
but most routines can also deal with symbolic powers. Similar notation is used in many other software packages related to multiloop calculations, such as \textsc{FIRE}, \textsc{LiteRed},  \textsc{KIRA} or \textsc{pySecDec}.

The function \texttt{FCFeynmanPrepare} is used to derive the Symanzik polynomials $\mathcal{U}$ and $\mathcal{F}$ for the given topology or set of master integrals. It can be invoked, not only on \texttt{FCTopology} or \texttt{GLI} objects, but also on integrals using explicit propagator representation via \texttt{Feyn\-Amp\-Denominator}. The underlying algorithm is based on the code \textsc{UF.m}~\cite{book:Smirnov:2012gma} that is used in \textsc{FIRE} and \textsc{FIESTA} \cite{Smirnov:2015mct,Smirnov:2021rhf}. The $\mathcal{U}$ and $\mathcal{F}$ polynomials encode numerous properties of the related topologies or master integrals  (\cf ref.~\cite{Bogner:2010kv} for an extensive review) and can be used to derive one-to-one mappings between those objects.

\texttt{FCFeynmanPrepare} is not limited to the derivation of Symanzik polynomials. It can also calculate other useful quantities such as the matrix $M$ with $\mathcal{U} = \det M$ or $J$ and $Q^\mu$ as in $\mathcal{F} = \det M (Q M^{-1} Q - J)$. Moreover, the routine is capable of dealing with both Minkowskian and Euclidean integrals. To avoid any confusion, let us stress that with ``Euclidean'' we explicitly mean integrals defined in the flat space with the Euclidean metric signature $g_E^{\mu \nu} = \text{diag} (1,1,1,1)$. To that aim one needs to set the option \texttt{"Euc\-lidean"} to \texttt{True}.

\subsection{Basic operations}

The ``old'' loop-related \textsc{FeynCalc} functions such as \texttt{TID}, \texttt{FDS} or \texttt{PaVe\-Reduce} are designed to work with loop integrals in the propagator representation. Therefore, upon introducing the new \texttt{GLI}-representation it became necessary to add a large set of new routines that accept input containing \texttt{GLI}- and \texttt{FCTopology}-symbols. However, in some cases the existing routines were just modified to be able to deal with the new objects.

One of the simplest manipulations applicable to a \texttt{GLI} is the conversion into the propagator representation. This can be done using \texttt{FCLoopFromGLI}. This function requires two arguments, which are a \texttt{GLI} integral and the corresponding topology in the form of an \texttt{FCTopology}. Both arguments can be also lists, which is useful when processing multiple integrals in one go.

Since a topology has a rather involved syntax, it can be validated using \texttt{FCLoop\-Valid\-TopologyQ}. This helps to avoid user errors when entering topologies by hand or converting them from the output of other tools. A list of all kinematic invariants present in a topology (or a list thereof) can be obtained with the aid of \texttt{FCLoopGetKinematicInvariants}.

A priori, the set of propagators contained in an \texttt{FCTopology} does not necessarily have to form a basis. However,  since many loop-related manipulations make sense only when working with a proper propagator basis, \textsc{FeynCalc} provides tools to verify this property. These are \texttt{FCLoopBasis\-OverdeterminedQ} and \texttt{FCLoopBasisIncompleteQ}, which can tell whether the given set of propagators is overdetermined or incomplete. In the latter case the basis can be automatically completed with suitable propagators using \texttt{FCLoop\-Basis\-Find\-Completion}. 

Integrals containing too many propagators must be subjected to partial fraction decomposition. Although \textsc{FeynCalc}'s \texttt{ApartFF} can now handle \texttt{GLI}s, in the context of loop calculations done with \texttt{FORM}, one would usually like to get explicit replacement rules that rewrite a product of overdetermined denominators into a linear combination of terms with fewer denominators. To cover that case we introduced a new routine called \texttt{FCLoop\-Create\-Partial\-FractioningRules} that will generate such rules and return a list of new topologies appearing on the right-hand side of the replacement rule.

Sometimes one might be interested in selecting particular topologies from a large list under the condition that those appear in the given loop integrals. To this aim we can use \texttt{FCLoopSelectTopology}, which is also employed by numerous high-level \textsc{FeynCalc} functions.

Differentiation of loop integrals with respect to vectors (similar to what can be achieved with \texttt{FourDivergence}) or scalars is implemented in \texttt{FCLoop\-GLI\-Differentiate}. This can be used \eg when deriving symbolic IBP relations, systems of differential equations or performing asymptotic expansions \cite{Beneke:1997zp}. The routine is therefore very similar to \textsc{LiteRed}'s \texttt{Dinv}. Notice that to have a proper differential equation, the differentiated loop integrals still require an IBP reduction. In this sense the function does not give one the final equation right away but merely performs one important step in the derivation thereof. Equally, it does not implement any methods to solve the resulting equation.

When doing asymptotic expansions one normally would like to attach a particular scaling parameter (say $\lambda$) to specific masses or momenta in the topology and expand the loop integrals in $\lambda$ to the given order. The former can be accomplished via \texttt{FCLoopAddScalingParameter} while for the latter one would use \texttt{FCLoopGLIExpand}.

An important point to keep in mind when working with loop integrals is the  $i \eta$-prescription in the propagators. By default, \textsc{FeynCalc} uses the standard convention, where a Minkowskian propagator is understood to be
\begin{align}
	[p^2 - m^2 + i \eta]^{-1} \label{eq:ieta}
\end{align}
However, an alternative prescription used \eg in \textsc{FIESTA}  is to pull out an overall minus sign, which leads to
\begin{align}
	[- p^2 + m^2 - i \eta]^{-1} \label{eq:ieta2}
\end{align}
Notice that this propagator is still Minkowskian, just written in different way as compared to Eq.~\eqref{eq:ieta}. For the sake of completeness, we list the relevant conventions for \textsc{FeynCalc}, \textsc{FIESTA} and \textsc{pySecDec} in Table~\ref{tab:eta-signs}. Notice that when using \texttt{SFAD} and \texttt{GFAD} shortcuts to enter loop integral propagators, \textsc{FeynCalc} will explicitly display $i \eta$, unless the global variable \texttt{\$FCShowIEta} has been set to \texttt{\$False}.
{
	\footnotesize 
	\renewcommand{\arraystretch}{1.5}
	\begin{longtable}[b]{|c|c|c|}		
		\hline
		Symbolic expression& Meaning in \textsc{FIESTA}   & Meaning in \textsc{FC} / \textsc{pySecDec}  \\
		\hline
		$[(p+q)^2]^{-1}$ & ${\color{red}[(p+q)^2 - i \eta]^{-1}}$ & ${\color{black!60!green}[(p+q)^2 + i \eta]^{-1}}$ \\
		\hline
		$[-(p+q)^2]^{-1}$ & ${\color{black!60!green}[-(p+q)^2 - i \eta]^{-1}}$ & ${\color{red}[-(p+q)^2 + i \eta]^{-1}}$  \\
		\hline
		$[(p+q)^2 - m^2]^{-1}$ & ${\color{red}[(p+q)^2 -m^2 - i \eta]^{-1}}$ & ${\color{black!60!green}[(p+q)^2 -m^2 + i \eta]^{-1}}$  \\
		\hline
		$[-(p+q)^2 + m^2]^{-1}$ &  ${\color{black!60!green}[-(p+q)^2 + m^2 - i \eta]^{-1}}$ & ${\color{red}[-(p+q)^2 + m^2 + i \eta]^{-1}}$  \\
		\hline
		\caption{Differences between different computer codes in the assumed $i \eta$-prescription for propagators. Expressions in green correspond to input yielding a correct imaginary part, while red terms will create
			inconsistencies.}
		\label{tab:eta-signs}
	\end{longtable}
}

\textsc{FeynCalc} can convert topologies to the convention of Eq.\eqref{eq:ieta2} via the function \texttt{FCLoop\-Switch\-Eta\-Signs}. \texttt{FC\-Loop\-Get\-Eta\-Signs} is used internally to ensure the consistency of the $i \eta$-prescription among propagators present in integrals and topologies.

\subsection{Topology identification}

Given a multiloop amplitude expressed as a linear combination of scalar loop integrals, one usually wants to reduce these integrals to a basis of master integrals. This procedure, usually referred to as the IBP-reduction, is not a mere convenience, but a strict necessity. While the number of unreduced loop integrals can easily go into hundreds of thousands or even millions, the number of master integrals often lies between $\mathcal{O}(100)$ and $\mathcal{O}(1000)$ for two- and three-loop calculations that are feasible with modern techniques. In practice, the reduction often turns out to be one of the main bottlenecks in analytic calculations and it is imperative to organize it in the most efficient way. For example, reducing each loop integral separately would be a waste of computational resources that should be avoided.

A better approach is to organize integrals into families and then do the reduction for each family. An integral family or a topology is defined as a set of linear independent propagators plus additional kinematic constraints such as values of masses or external momenta squared. It goes without saying that the number of families should better be as small as possible, otherwise one would be wasting computer time and resources. A caveat, however, lies in the fact that dimensionally regularized loop integrals are invariant under shifts of loop momenta. Hence, two integrals that look very different might still belong to the same family. Also, two topologies that seem to be quite distinct could represent the same quantity modulo momentum shifts.

Such ambiguities can be avoided using  a procedure called topology identification or minimization, where the set of all loop integral topologies present in the amplitude is mapped to a smaller set of topologies independent of each other. Most algorithms for solving this task either consider graph representations of Feynman diagrams and amplitudes or analyze the propagators present in the amplitude. In the former case the initial problem of finding mappings between different topologies is converted into the requirement to find subgraph isomorphisms. When working with symbolic propagators one is mainly interested in finding a representation that removes the invariance under momentum shifts and generates unique expressions that can be directly compared with each other. Of course, a brute-force enumeration of all possible momentum shifts is also possible, although mostly prohibitively expensive performance-wise. A purely graph-based algorithm is implemented \eg in the \textsc{C++} programs \textsc{q2e/exp}, while the \mma package \textsc{TopoID}~\cite{Hoff:2016pot} and the generator of optimized IBP identities
\textsc{NeatIBP}~\cite{Wu:2023upw} look only at propagators of the loop integrals. Hybrid approaches are realized \eg in \textsc{Reduze}, \textsc{feynson}, \textsc{tapir} or \textsc{pySecDec}.

\textsc{FeynCalc} follows a purely propagator-based approached by using the so-called Pak algorithm~\cite{Pak:2011xt} --- a special prescription for comparing topologies or integrals with each other invented by Alexey Pak. Our implementation heavily relies on the ideas and tricks that can be found in the doctoral thesis of Jens Hoff~\cite{Hoff:2015kub} and their realization in Hoff's program \textsc{TopoID}\footnote{\url{https://github.com/thejensemann/TopoID}}.

The starting point for Pak's algorithm is the naive observation that in the Feynman parametric representation of loop integrals (or topologies) the loop momenta are integrated out. Hence, the shift invariance seems to be gone. Unfortunately, there still remains a residual ambiguity, which is related to the relabeling of Feynman parameters $x_i \leftrightarrow x_j$. Pak's insight was to introduce a canonical way to label $x_i$ for the given combination of $\mathcal{U}$ and $\mathcal{F}$ polynomials. Then, given two canonically ordered characteristic polynomials $\mathcal{P}_1 \equiv \mathcal{U}_1 \times \mathcal{F}_1$ and $\mathcal{P}_2 \equiv \mathcal{U}_2 \times \mathcal{F}_2$, it is guaranteed that for identical integrals or topologies we will find  $\mathcal{P}_1 = \mathcal{P}_2$. This property is the corner stone for \textsc{FeynCalc}'s functionality of finding one-to-one mappings between topologies or master integrals.

When implementing this technique we introduced a number of auxiliary functions that return the necessary building blocks for applying Pak's algorithm. For example, \texttt{FCLoopToPakForm} can be used to generate the canonically ordered characteristic polynomial $\mathcal{P}$ from the given propagator representation, while \texttt{FCLoopPakOrder} can apply Pak ordering to any polynomial.

A nice property of $\mathcal{P}$ is that it can be used to detect scaleless integrals that
vanish in dimensional regularization. The description of the underlying algorithm can be 
found in Section~2.3 of ref.~\cite{Hoff:2016pot}. In \textsc{FeynCalc} the scalefulness property can be checked using the functions
\texttt{FCLoopPakScalelessQ} (for characteristic polynomials) and \texttt{FCLoopScalelessQ} (for loop integrals).

The workflow envisioned in \textsc{FeynCalc} begins with applying \texttt{FCLoop\-Find\-Topo\-logies} to the given amplitude. The function will return a list of the form \texttt{\{amp, topos\}}, with \texttt{amp} being the amplitude rewritten in a form suitable for further processing and \texttt{topos} constituting a list of all distinct sets of propagator denominators. In \texttt{amp} these sets are grouped into \texttt{GLI}s, while \texttt{topos} is made of \texttt{FCTopology} objects. In the next step one should get rid of tensor integrals, unless this has already been done by applying suitable projectors.

To this aim one can use the function \texttt{FCLoopTensorReduce} --- which still uses 
\texttt{Tdec} as back-end, but is optimized for the representation of the amplitude generated 
by \texttt{FCLoop\-Identify\-Topologies}. Alternatively, one could also apply \texttt{FCMultiLoopTID} to the amplitude before running \texttt{FCLoop\-Identify\-Topologies}, but for performance reasons we do not recommend this. It is also worth mentioning that the old code used in \texttt{Tdec}\footnote{\texttt{Tdec} is an auxiliary routine generating tensor decomposition formulas for generic multiloop integrals that can be used in \textsc{FeynCalc} or \textsc{FORM} \cite{Reeck:2024iwk}.} for recognizing symmetries between tensor reduction coefficients has been replaced with the algorithm described in ref.~\cite{Pak:2011xt}. On selected examples the new symmetrizer can lead to systems of linear equations being much smaller as compared to \textsc{FeynCalc} 9.3.1. Still, for higher rank tensor integrals \mma's capabilities might be insufficient to solve the linear system in a reasonable amount of time. This issue can be worked around using the new version of \textsc{FeynHelpers}, which allows \texttt{Tdec} to use \textsc{FERMAT}~\cite{Lewis:Fermat} as a solver back-end.

The one-to-one mappings between topologies can be revealed by applying  \texttt{FCLoop\-Find\-Topology\-Mappings} to the \texttt{topos} list. Every mapping rule between two topologies contains of a list of momentum shifts that convert the propagators of the first topology into those of the second topology and should  be  also applied to all numerators multiplying the first topology. It is also possible to map the given topologies onto a specific set of selected
topologies (\eg to facilitate comparisons to other calculations) using the option
\texttt{PreferredTopologies}.

Due to the nature of Pak's algorithm \texttt{FCLoopFindTopologyMappings} can only find relations between topologies that contain exactly the same number of propagators. Since not all topologies appearing in the amplitude normally have the full set of propagators required to form a basis, this often leaves some room for mapping smaller topologies into larger ones. 
Here with ``larger topologies'' we mean both incomplete topologies with a larger number of propagators as well as the so-called supertopologies that have a complete propagator basis. In \textsc{FeynCalc} one can deal with this situation by first identifying all nonvanishing subtopologies of the given topology via \texttt{FCLoopFindSubtopologies}. Then, one can try to find mappings between those subtopologies and the actual smaller topologies appearing in the amplitude. The subtopologies contain a special marker that relates them to the parent topology, so that \texttt{FCLoopFindTopologyMappings} knows how to generate correct mappings pointing to the original topology.

Once the final set of topologies has been sufficiently minimized, one can apply the generated mapping rules to the full amplitude with the aid of \texttt{FCLoopApplyTopologyMappings}. In fact, this routine will also rewrite all scalar products in terms of invert propagators and combine them with the existing propagator denominators, so that the resulting amplitude will appear as a linear combination of different \texttt{GLI}s. In the background this high-level function uses the auxiliary routines \texttt{FCLoopCreateRulesToGLI} and \texttt{FCLoop\-Create\-Rule\-GLI\-To\-GLI}. The output of \texttt{FCLoopApplyTopologyMappings} is, in principle, suitable for the subsequent IBP reduction. The process of converting the occurring loop integrals and the list of final topologies into run cards for tools such as \textsc{FIRE} and \textsc{KIRA} can be easily automatized. The relevant code is already part of \textsc{FeynHelpers} that will be presented elsewhere.

\section{Master integrals} \label{sec:masters}

Unless one is trying to perform a cutting-edge calculation\footnote{Of course, if the given calculation involves enough masses and/or legs, even the reduction of two-loop amplitudes can quickly become unfeasible.} or lacks access to sufficient computational resources, the IBP reduction usually goes through without much additional effort. Even when some fine-tuning is needed, this usually amounts to playing with the configuration files of the respective programs, resubmitting jobs running on a cluster or possibly asking the developers for an advice. Once all reduction tasks have been completed, one is normally left with a list of master integrals from different integral topologies.

Depending on the tool that was used to perform the reduction, it may be necessary to check whether all of these integrals are indeed distinct. Given that identical masters can have rather different propagator representations,
this task should be better performed in an algorithmic fashion. \textsc{FeynCalc} can make use of the built-in Pak's method to reveal all one-to-one mappings between the master integrals. The corresponding function is called \texttt{FCLoopFindIntegralMappings} and has been modeled after the routine \texttt{FindRules} in \textsc{FIRE}. Using the option \texttt{PreferredIntegrals} one can choose a list of preferred master integrals to be mapped onto. This also works for factorizing integrals, which can be entered as products of \texttt{GLI}s.

Graphical representations of master integrals serve as an important tool to better understand the obtained results and relate them to the calculations that has already been done in the literature. The most common visualization method is to relate the propagators and the flow of loop momenta inside the integral to graphs made of directed edges. These edges can be styled
to account for the types of propagators and their masses. For example, massless propagators are usually plotted as dashed or dotted lines, while massive propagators are shown as solid lines of different colors. If the propagator appears squared, this can be hinted using a dot or a cross.

While switching from a graph to a propagator representation for the given integral (or topology) is a trivial step, the converse is not true. The construction of graphs for arbitrary integrals can be tricky and requires both care and effort. Some publicly available tools such as \textsc{AZURITE}, \textsc{PlanarityTest}~\cite{Bielas:2013rja} or \textsc{LiteRed} allow for automatizing this task to some extent. In \textsc{FeynCalc}, the corresponding function is called \texttt{FCLoopIntegralToGraph}. In the case of a successful reconstruction, it returns a directed \mma graph as well as 
the line momenta running through the edges and some additional information. Although the graph can be directly shown using the built-in \mma function \texttt{GraphPlot}, one should better use \textsc{FeynCalc}'s \texttt{FCLoopGraphPlot} which takes options for styling the edges and making the output look more similar to what is usually seen in the literature. However, it should be noted that due to multiple problems that older \mma versions have with the visualizations of graphs, \texttt{FCLoopGraphPlot} requires at least \mma 11.0, while best results can be expected with the version 12.2 or newer. We would also like to stress that the graph obtained with \texttt{FCLoopIntegralToGraph} can also (upon some minimal adjustments) be plotted using other suitable software such as \textsc{GraphViz}.

One of the advantages of having master integrals shown as graphs is that one can readily assess possible cuts via visual examination. In this context we understand ``cutting'' as the process of sending the cut propagators on-shell so that pictorially the graph splits into two graphs. However, analyzing dozens or even hundreds of graphs by eye can still be tedious and prone to human errors. To streamline this process \textsc{FeynCalc} offers \texttt{FCGraphCuttableQ}, which can decide whether the given graph can be cut without touching the specified lines. This is relevant \eg for heavy particles in the loops that kinematically cannot go on-shell. This way one can readily determine whether the given master integrals can develop an imaginary part or not. A more generic routine is offered under the name \texttt{FCGraphFindPath}. Its task is to determine,  whether the given  graph can be traversed by starting and finishing at one of the external edges. The internal edges can be assigned weights $1$ or $-1$, with the latter meaning that  this edge cannot be passed.

Having dealt with the problem of obtaining a set of unique master integrals and visualizing them, it is fair to ask whether \textsc{FeynCalc} can also be useful for evaluating the master integrals. As far as numerical evaluation is concerned, the \textsc{FeynHelpers} interface (to be presented in a future publication) makes it easy to generate ready-to-use \textsc{FIESTA} or \textsc{pySecDec} scripts for evaluating the given \texttt{GLI}s. Analytic results are of course  much more difficult to obtain. Calculating loop integrals in this fashion often involves trying or even combining different techniques available on the market (\cf \eg refs.~\cite{Smirnov:2006ry,Weinzierl:2022eaz}) in the hope that an integral, that is intractable using method A may turn out to be easy when attacked with method B. While \textsc{FeynCalc} obviously cannot deliver analytic solutions upon pressing a button, it nevertheless offers a set of handy routines that facilitate common steps required for some calculational methods.

As far as the derivation of the Symanzik polynomials $\mathcal{U}$ and $\mathcal{F}$ is concerned, the previously mentioned function \texttt{FCFeynmanPrepare} can readily generate those expressions. If one is interested in evaluating the master integral via a direct analytic integration of its Feynman parametric representation, the more useful routine would be \texttt{FCFeynmanParametrize}. Notice that this function supports both quadratic and eikonal propagators and can also deal with Euclidean or tensor integrals. Cartesian integrals living in $D-1$ dimensions are equally supported.

In general, it is very difficult to carry out all Feynman parametric integrations while keeping
the full dependence on the dimensional regulator $\varepsilon$. For simpler integrals this
can be often achieved by first joining specific subsets of propagators before combining the rest
in the final integrand. This trick may often result in a greater freedom when exploiting the
Cheng--Wu theorem~\cite{Cheng:1987ga} and trying to find a working sequence of integrations. To this aim
\textsc{FeynCalc} offers \texttt{FCFeynmanParameterJoin}, which makes the unification of different propagators
simple and straightforward. Its output can be then passed to \texttt{FCFeynmanParametrize}, thus
obtaining the final integrand depending on the introduced sets of Feynman parameters $x_i$, $y_i$, $z_i$ etc.

The applicability of the Cheng--Wu theorem can be readily checked via \texttt{FCFeynman\-Pro\-jective\-Q}.
If, for some reason, the integral turns out not to be projective, it can be rendered projective using 
\texttt{FCFeynmanProjectivize} by automatically performing a projective transformation. 

If one insists on integrating Feynman parameters order by order in $\varepsilon$, one should keep in mind that naively expanding the integrand in the dimensional regulator may introduce divergences in the Feynman parameters, which is clearly undesirable. One possible solution to this problem involves the so-called analytic regularization that was
developed in refs.~\cite{Panzer:2014caa,Panzer:2014gra,Panzer:2015ida} and implemented in the \textsc{Maple} package \textsc{HyperInt}~\cite{Panzer:2014caa}. In a nutshell, when expanding an integrand that has been treated using this technique, all $\varepsilon$ poles become explicit so that the integrations in $x_i$ remain finite. Unlike sector decomposition, analytic regularization is guaranteed not to spoil the linear reducibility property of the 
integrand. In \textsc{FeynCalc} analytic regularization is implemented using the functions \texttt{FC\-Feynman\-Find\-Divergences} and \texttt{FCFeynmanRegularizeDivergence}, that were inspired by \textsc{HyperInt}'s \texttt{findDivergence} and \texttt{dimregPartial}.

Nowadays, the method of differential equations~\cite{Kotikov:1991pm,Kotikov:1990kg,Kotikov:1991hm,Bern:1993kr,Remiddi:1997ny,Gehrmann:1999as} belongs to the most popular and efficient techniques for calculating large numbers of master integrals analytically or numerically. The discovery of the canonical form \cite{Henn:2013pwa,Henn:2014qga} and a rapid advance in the software for automatically finding such forms was very beneficial for the field of multiloop calculations. When using tools such as  \textsc{Fuchsia}~\cite{Gituliar:2017vzm}, \textsc{CANONICA}~\cite{Meyer:2016slj,Meyer:2017joq}, \textsc{Libra}~\cite{Lee:2014ioa,Lee:2020zfb}, \textsc{epsilon}~\cite{Prausa:2017ltv}, \textsc{Initial}~\cite{Dlapa:2020cwj,Dlapa:2022wdu} etc. one is often confronted with the necessity to perform a change of variables \eg for rationalizing square roots appearing at intermediate stages\footnote{The transformations intended to remove such square roots can be automatically obtained using \textsc{RationalizeRoots} \cite{Besier:2019kco}}. To automatize this step \textsc{FeynCalc} offers a function called \texttt{FCDiffEqChangeVariables}. At the moment only differential equations
of one variable are supported. Given the old variable $x$, the new variable $y(x)$ as well as the inverse relation $x(y)$, this routine eliminates $x$ in favor of $y$ in the given matrix. This can be $\mathcal{A}$ from the pre-canonical form of the differential equation $F' =\mathcal{A} F$, but also $\mathcal{B}$ from the canonical form $G' = \varepsilon \mathcal{B} G$ or just the transformation matrix $\mathcal{T}$ with $F = \mathcal{T} G$. Notice that in the case of $\mathcal{T}$ one should disable the inclusion of the prefactor $1/f'(y)$ to avoid incorrect results. This is done by setting the option \texttt{Prefactor} to \texttt{False}.

Having obtained a canonical form using one of the existing tools, one usually starts constructing the solution to the system order by order in $\varepsilon$. Here one can make use of \texttt{FCDiffEqSolve} that can quickly generate such expressions written in terms of \texttt{FC\-Iterated\-Integral} objects. The latter can be regarded as a placeholder for expressions of the type
\begin{equation}
	\int_a^b d x f(x),
\end{equation}
where $f(x)$ can be an iterated integral itself. This way nested integrals can be represented in \textsc{FeynCalc} by wrapping new \texttt{FCIteratedIntegral} heads around the existing ones.

To facilitate the evaluation  of such integrals, the rational functions involved are transformed into a special representation called \texttt{FCPartial\-Fraction\-Form}. The main idea is to write expressions of the form 
\begin{equation}
	n + \frac{f_1}{[x-r_1]^{p_1}} + \frac{f_2}{[x-r_2]^{p_2}} + \ldots
\end{equation}
as \texttt{FCPartialFractionForm[n,\{\{f1,x-r1,p1\},\{f2,x-r2,p2\},...\},\-x]}. From here one can easily rewrite \texttt{FCIteratedIntegral} objects in terms of Harmonic or Goncharov polylogarithms \cite{Remiddi:1999ew,Goncharov:1998kja,Goncharov:2001iea}. This is done using \texttt{FCIteratedIntegralEvaluate} with the result containing \texttt{FC\-GPL} symbols. The conversion of rational functions to this representation is handled by \texttt{To\-FC\-Partial\-FractionForm}.

Notice that for the time being, \texttt{FCGPL}s are mere placeholders. More GPL-related routines are expected to be added  in future versions of \textsc{FeynCalc}. Our goal is to have some minimal implementation of core symbolic properties of GPLs that can be used by the related \textsc{FeynCalc} functions (mainly for computing loop integrals) without the need to employ any external packages. This way we can ensure that there will not be any unwanted side effects that often arise when having multiple packages loaded on the same kernel or when using packages that have not yet been made 100\% compatible with the most recent \textsc{Mathematica} version\footnote{For example, \textsc{PolyLogTools} currently relies on the \textsc{Combinatorica} library that is being deprecated since \textsc{Mathematica} 10 and generates multiple warnings when loaded on version 13.3 or 14.0. The \textsc{HPL} package that is equally required by \textsc{PolyLogTools} is not being actively maintained since years and gets patched on the fly whenever it is loaded by \textsc{PolyLogTools}. Then, \textsc{MPL} is a \textsc{Maple} package and hence cannot be used directly with \textsc{Mathematica}.}.

Of course, for extensive manipulations of multiple polylogarithms the users should resort to special codes
such as \textsc{HPL} \cite{Maitre:2005uu,Maitre:2007kp}, \textsc{PolyLogTools}~\cite{Duhr:2019tlz} or \textsc{MPL}~\cite{Bogner:2015nda} that have been developed over the years and are well established in the field.

\section{Features and improvements unrelated to multiloop calculations} \label{sec:rest}

\subsection{Improved color algebra simplifications}
In the past, \textsc{FeynCalc} was often unable to simplify various $\text{SU}(N)$ color algebraic expressions using \texttt{SUNSimplify} and \texttt{SUNTrace}. Sometimes chaining multiple instances of the two routines with different options would do the trick, but such workarounds were different to find and far from being obvious to users.

To improve on this situation, in \textsc{FeynCalc} 10 the function \texttt{SUNSimplify} was rewritten from scratch. The new version implements a much larger number of color algebraic relations, while the new code is easier to maintain and extend. Notice that the evaluation of the color trace is now handled in a manner similar to what is done in \texttt{DiracSimplify}. By default, an \texttt{SUNTrace} object remains unevaluated, unless the option \texttt{SUNTraceEvaluate} is set to \texttt{True}. However, the more convenient way to evaluate such expressions is to use \texttt{SUNSimplify}. The default value of the \texttt{SUNTraceEvaluate} option in \texttt{SUNSimplify} is set to \texttt{Automatic}. This means that if a trace can be simplified without naively rewriting everything in terms of structure constants, the function will do so. Setting this option to \texttt{False} will leave all traces untouched, while \texttt{True} means that the user explicitly wants to eliminate the traces in favor of \texttt{SUNF} and \texttt{SUND} symbols.

\subsection{Passarino--Veltman functions}

As a package, that was originally developed with one-loop calculations in mind, \textsc{FeynCalc} is of course equipped with symbols representing Passarino--Veltman functions and a set of routines for working with them. One particular shortcoming related to this functionality that became obvious in the past few years, was \textsc{FeynCalc}'s ignorance of many symmetry relations between \texttt{PaVe} functions. This way some results looked longer and more complicated than they actually should have been and certain cancellations did not take place.

In \textsc{FeynCalc} 10 we tried to add all symmetries up to rank 10 for $B$-functions, rank 9 for $C$-functions, rank 8 for $D$-functions, rank 7 for $E$-functions and rank 6 for $F$-functions. The corresponding files are located inside the directory \texttt{Tables/PaVeSymmetries} and can be (if needed) extended to even higher ranks. Whenever the user enters \texttt{PaVe} functions, \texttt{PaVeOrder} will automatically reorder their arguments in a canonical way, unless the option \texttt{PaVeAutoOrder} has been explicitly set to \texttt{False}.

Also, the deprecated \texttt{OneLoop} routine offered a functionality that was tricky to reproduce using other functions: The ability to simplify IR-finite expressions involving \texttt{PaVe} functions by analyzing their UV-poles and expanding the $D$-dependent prefactors accordingly so that the expression becomes $\mathcal{O}(\varepsilon^0)$. To improve on this, we added \texttt{PaVeLimitTo4} which does exactly that. Notice that the absence of IR-poles is assumed but not explicitly checked, meaning that it is the user's duty to ensure this condition's validity.

\subsection{Lagrangians and operators}

In the course of our ongoing work to improve the usefulness of \textsc{FeynCalc} for nonrelativistic calculations, we extended the functionality of the package for manipulating Lagrangians to support Cartesian nabla operators. Supplementing the already existing symbols \texttt{LeftPartialD} ($\sim \overleftarrow{D}^\mu$), \texttt{RightPartialD} ($\sim \overrightarrow{D}^\mu$), \texttt{LeftRightPartialD} ($\sim \overleftrightarrow{D}^\mu$) and \texttt{LeftRight\-Partial\-D2} ($\sim \overleftrightarrow{D}^2$) we now also have \texttt{LeftNablaD} ($\sim \overleftarrow{\nabla}^i$), \texttt{RightNablaD} ($\sim \overrightarrow{\nabla}^\mu$), \texttt{LeftRightNablaD} ($\sim \overleftrightarrow{\nabla}^\mu$) and \texttt{LeftRightNablaD2} ($\sim \overleftrightarrow{\nabla}^2$).

Notice that although one still cannot use \texttt{FeynRule} to derive Feynman rules for nonrelativistic operators, other useful routines such as \texttt{Expand\-Partial\-D} and \texttt{Explicit\-Partial\-D} can now deal with nabla operators or gauge covariant derivatives with Cartesian indices (\ie $D^i$ from $D^\mu=(D^0,D^i)$).

Another useful addition to this part of \textsc{FeynCalc}'s capabilities is \texttt{Shift\-Partial\-D}, which allows the user to reshuffle derivatives in specific operators by applying integration by parts on the Lagrangian level. In this case the surface terms are always assumed to vanish.

Last but not least, in order to further facilitate the process of writing custom functions working with \textsc{FeynCalc} symbols (\eg for deriving Feynman rules in the spirit of Appendix C from \cite{Brambilla:2020fla}), version 10 also features \texttt{FCTripleProduct} as a shortcut for vector products $(\vec{a} \times \vec{b}) \cdot \vec{c}$ as well as two routines for extracting all free or dummy  indices in the given expression. They are called \texttt{FCGetFreeIndices}  and \texttt{FCGetDummyIndices} respectively.

\subsection{Dirac algebra}

It is now possible to apply Gordon identities to suitable spinor chains by means of \texttt{GordonSimplify}. The function works both in 4 and $D$ dimensions, while the option \texttt{Select} allows to choose whether one wants to trade the right-handed projector $P_R$ for the left-handed $P_L$ one or vice versa.

Furthermore, the calculation of Dirac traces in the Larin~\cite{Larin:1993tq} scheme now proceeds according to the so-called Moch-Vermaseren-Vogt \cite{Moch:2015usa} formula, which greatly improves the computational efficiency as compared to the previous implementation.

The code for the evaluation of some special spinor chains such as $\bar{v}(p) \gamma^5 v(p)$, $\bar{u}(p) u(p)$ or
$\bar{v}(p) v(p)$ was moved from \texttt{Dirac\-Simplify} to a dedicated routine called \texttt{Spinor\-Chain\-Evaluate}. Setting the same-named option of \texttt{Dirac\-Simplify} to \texttt{False} will prevent \textsc{FeynCalc} from  replacing such objects with their explicit values --- which can be useful for certain types of calculations.

Unfortunately, \textsc{FeynCalc} still does not support the spinor-helicity formalism~\cite{Berends:1981rb,Caffo:1982ds,DeCausmaecker:1981jtq,Xu:1984qe,Kleiss:1985yh,Gunion:1985vca,Xu:1986xb}, which constitutes a much more efficient way to deal with fermions, especially in the massless case. This feature remains on our to-do list.

\subsection{Convenience functions for research activities}

Some of the functions introduced in \textsc{FeynCalc} 10 are not directly related to the evaluation of amplitudes or loop integrals but rather belong to the category of the so-called convenience routines. One of them is called \texttt{FCMatchSolve} and has been developed to automatize the determination of renormalization constants, matching coefficients and other parameters. To this aim, for a given expression (\eg the difference of two amplitudes or the sum of some diagrams and the corresponding counterterms), one first needs to collect all unique structures (\eg Dirac chains, color factors, $\varepsilon$, $\alpha_s$ etc.). Then, one can pass this expression to \texttt{FCMatchSolve} together with the list of symbols that should be regarded as fixed variables. In this case the function regards all other variables as free parameters and tries to choose them such, that the input expression vanishes. In practice, this approach turns out to be more efficient and robust than using \texttt{Collect} and \texttt{Solve}.

Another common task in particle phenomenology is the numerical evaluation of the final analytic expressions for cross sections, decay rates, matching coefficients and other experimentally accessible parameters. Comparing such quantities to the literature or to the results of peers can be nontrivial for several reasons: Firstly, contrary to symbolic expressions, the comparison will not be exact, but rather up to a given number of $n$ significant digits. Second, when finding disagreement between two large expressions involving numbers of different origin, one would often want to identify terms that agree with less significant digits than required, rather than merely state the lack of numerical agreement. Using \texttt{FCCompareNumbers} one can streamline the task of comparing two numerical or semi-numerical expressions, while retaining full control over the number of significant digits required. Again, even though a similar result could be achieved using custom codes, \texttt{FCCompareNumbers} is an attempt at saving time by automating trivial operations and avoiding the most common pitfalls of manual evaluation.

Putting the often long and complicated analytic expressions obtained in a multiloop calculation into a proper form suitable for a publication, can be regarded as an art of its own. When using \mma for organizing the expressions and converting them into \LaTeX, one is often faced with the problem that sums of terms are not ordered in the way one would want them to. This is because \mma's \texttt{Times} and \texttt{Plus} functions sort terms using an internal canonical ordering that does not necessarily agree with one's aesthetic preferences. To remedy this, \textsc{FeynCalc} now comes with a function called \texttt{FCToTeXReorder} that first converts \texttt{Times}- and \texttt{Plus}-type expressions into nested lists of the form \texttt{\{a,b,...,Plus\}}
and \texttt{\{a,b,...,Times\}} respectively. Terms inside those lists can then be grouped and ordered according to the
user's preferences, using custom factoring and sorting functions. The intermediate result of such manipulations
can be readily previewed with \texttt{FCToTeXPreviewTermOrder}. Once the expressions have been brought into a suitable
form, one can directly apply the built-in \texttt{TeXForm} command to the output of \texttt{FCToTeXPreviewTermOrder} and
then copy the generated \LaTeX \,code into the source file of the publication.

\subsection{Tensors with light-cone components} \label{sec:lc}
In many QFT calculations (especially those involving highly energetic particles) it is natural
to decompose Lorentz tensors into components along two light-like reference vectors $n$ and $\bar{n}$ satisfying
\begin{equation}
	n^2 = \bar{n}^2 = 0, \quad n \cdot \bar{n} = 2 \label{eq:lcvecs}.
\end{equation}
For example, a four-vector can be then written as

\begin{equation}
	p^\mu = \frac{\bar{n}^\mu}{2} (p \cdot n)  + \frac{n^\mu}{2} (p \cdot \bar{n}) + p_\perp^\mu \equiv p^\mu_+ + p^\mu_- + p_\perp^\mu,
\end{equation}
with the perpendicular component being defined as the difference between the full vector and the sum of the plus and minus components
\begin{equation}
	p_\perp^\mu \equiv p^\mu - p_+^\mu  - p_\perp^\mu = p^\mu - \frac{\bar{n}^\mu}{2} (p \cdot n)  - \frac{n^\mu}{2} (p \cdot \bar{n}).
\end{equation}
To facilitate such calculations using \textsc{FeynCalc}, version 10 of the package introduces
special symbols for defining light-like reference vectors as well as additional quantities specifying the plus, minus and perpendicular components of Lorentz tensors. First of all, one has to tell \textsc{FeynCalc}, which symbols represent $n$ and $\bar{n}$ by assigning the corresponding values to \texttt{\$FCDefaultLightConeVectorN} and \texttt{\$FCDefaultLightConeVectorNB}. In addition to that, one should
also implement the constraints from Eq.~\eqref{eq:lcvecs} \eg as in
{\small
	\begin{mmaCell}[indexed=false,moredefined={FCClearScalarProducts,ScalarProduct,n,nb}]{Code}
		FCClearScalarProducts[]
		ScalarProduct[n,n] = 0; ScalarProduct[nb,nb] = 0;
		ScalarProduct[n,nb] = 2;
	\end{mmaCell}
}
After these preliminary steps one can start using new shortcuts for the lightcone components such as 
\texttt{FVPL[p,$\mu$]} for $p^\mu_+$ or \texttt{SPLR[p,q]}, $(p \cdot q)_\perp$ etc. In the case of plus
and minus components, \textsc{FeynCalc} would insert explicit expressions constructed from the full tensor contracted
with $n$ and $\bar{n}$ vectors. Perpendicular components are represented
using the symbol \texttt{LightConePerpendicularComponent}, which takes a \texttt{LorentzIndex} or \texttt{Momentum}
as first argument and requires \texttt{Momentum[n]} and \texttt{Momentum[nb]} for the remaining two arguments.

\textsc{FeynCalc} can work with expressions involving light-cone components of vectors, metric tensors and scalar products in $4$ or $D$ dimensions. Dirac matrices can be also defined on the light-cone --- with both \texttt{DiracSimplify} and \texttt{DiracTrace} able to simplify the corresponding expressions.

\subsection{Up-to-date documentation using continuous integration} \label{sec:docu}

With this release we also address shortcomings of the \textsc{FeynCalc} documentation: The lack of a proper manual in the form of a PDF file, new functions  introduced but not documented, the absence of a novice-friendly tutorial. Technical issues forced us to rethink the whole concept behind the documentation of the package. We decided to stop maintaining the documentation sources in form of \mma notebooks in favor of switching to  text-based \texttt{.m} and markdown files. Using a modified version of J. Podkalicki's \mma to Markdown converter \textsc{M2MD}\footnote{\url{https://github.com/kubaPod/M2MD}} together with the \textsc{Pandoc} \cite{url:pandoc} document converter, we created a workflow, where \texttt{.m} and \texttt{.md} (markdown) files containing the whole documentation can be semi-automatically converted to HTML (for the online documentation) or \LaTeX\,(for the PDF manual). The \LaTeX-form of the manual is kept in a separate repository\footnote{\url{https://github.com/FeynCalc/feyncalc-manual}} and every change in the source files triggers  an update of the public PDF file that can be readily downloaded\footnote{\url{https://github.com/FeynCalc/feyncalc-manual/releases/tag/dev-manual}} by anyone.  This way, it is easy to keep both the online and PDF versions of the manual up to date without the need to update or modify their content  manually. Furthermore, we also took care to automatically synchronize the descriptions of \textsc{FeynCalc} symbols and functions in the documentation  to the texts shown when looking up their usage information (\eg as in \texttt{?FV} or \texttt{?TID}). We hope and believe that these changes will significantly improve the user experience of \textsc{FeynCalc} and make the program more accessible to new users.

\section{Examples} \label{sec:examples}

In this section we would like to draw reader's attention to several examples included with the package that make use of the new multiloop-related routines. Unlike old \textsc{FeynCalc} codes, where one-loop amplitudes were always expressed in terms of Passarino--Veltman functions, the calculations presented below are carried out in a different way, where the resulting amplitudes are written in terms of \texttt{GLI}s belonging to previously identified integral families in the \texttt{FCTopology} notation.

\subsection{Electron self-energy in massless QED at 2 loops}

We start by generating the 3 required 2-loop diagrams in QED using \textsc{FeynArts}. Upon setting the electron mass to zero, we apply \texttt{Dirac\-Simplify} to the amplitudes and then continue to the topology identification stage. Using \texttt{FC\-Loop\-Find\-Topologies} we find two distinct sets of propagators that can be fitted into two trial topogies. However, upon identifying all nonvanishing subtopologies by means of \texttt{FCLoop\-Find\-Subtopologies}, we can apply \texttt{FCLoop\-Find\-Topology\-Mappings} and map everything into one single topology. 

The next step is to carry out the tensor reduction (\texttt{FCLoopTensorReduce}) and then apply mappings between trial topologies to the diagrams. This can be conveniently handled by \texttt{FCLoopApplyTopologyMappings}.
In addition to the above steps this routine will also rewrite the scalar products involving loop momenta in terms of inverse denominators (in the \texttt{GLI}-notation) and bring the amplitudes into a form where they are expressed as linear combinations of various \texttt{GLI} integrals. 

We omit the technicalities related to the IBP reduction (which can be also automatized using \textsc{FeynHelpers}) and use an already available reduction table to reduce everything to 3 master integrals. Two of them are, however, identical --- which can be revealed via \texttt{FCLoopFindIntegralMappings}. Thus, we obtain the final result 
for the 2-loop electron self-energy in massless QED expressed in terms of 2 master integrals. Comparing this to the result given in Eq.~5.51 of ref.~\cite{Grozin:2005yg} we find full agreement, as expected.

\subsection{Photon self-energy in massless QED at 2 loops}

This example is almost identical to the previous one apart from the obvious fact that we need to generate
another set of diagrams. The calculation is carried out with full gauge dependence, even though $\xi$ cancels
in the final result. The obtained result given in terms of 2 master integrals can be compared
in Eq.~5.18 of ref.~\cite{Grozin:2005yg}.

\subsection{Gluon self-energy in massless QCD at 2 loops}

The gluon self-energy calculation shares many similarities with the previous examples but also requires some adjustments. First of all, due to the complexity of this calculation (18 diagrams) we use Feynman gauge. Then,
we employ Lorentz and color projectors to extract the scalar self-energy function $\Pi(p^2)$ directly. Apart from these technicalities, the main steps of the calculation closely follow those of the two previous examples. Here we choose to insert explicit expressions for the 2 master integrals and compare the so-obtained results at $\mathcal{O}(\varepsilon^0)$ with the sum of Eqs.~6.10-6.11 in ref.~\cite{Davydychev:1997vh}, finding complete agreement.

\subsection{Topology identification for $B_c \to \eta_c$ form factors at 2 loops}

This example demonstrates the usage of \textsc{FeynCalc} for the sole purpose of minimizing a set of topologies obtained from elsewhere (\eg in \textsc{FORM} calculation). We refer to refs.~\cite{Boer:2023tcs,Shtabovenko:2024aum} for more details on the underlying physics and the technical aspects of this calculation. The given set of 251 2-loop topologies with 3 external momenta and 3 scales contains not only quadratic but also eikonal as well as mixed quadratic-eikonal propagators (\cf \ref{sec:mixedprops}). We first deal with the mixed propagators by completing the square using the routine \texttt{FC\-Loop\-Replace\-Quadratic\-Eikonal\-Propagators}. Then we address topologies containing overdetermined sets of propagators by generating the corresponding partial fraction decomposition rules via \texttt{FC\-Loop\-CreatePartial\-Fractioning\-Rules}. Having uncovered all suitable topology mappings with the aid of \texttt{FC\-Loop\-Find\-Topology\-Mappings}, we add missing propagators needed to have a complete basis in each topology (\texttt{FCLoopBasisFindCompletion}) and generate rules for rewriting scalar products containing loop momenta in terms of inverse propagators (\texttt{FCLoopCreateRuleGLIToGLI})
 so that the full amplitude can be written solely in terms of \texttt{GLI}s. All these results can be readily converted into \textsc{FORM} \texttt{id}-statements and thus used in a \textsc{FORM}-based setup.

\section{Summary} \label{sec:summary}

\textsc{FeynCalc} 10 is a big step towards the goal of bringing multiloop calculations closer to the broad audience of interested phenomenologists. The presented release of the package integrates numerous new functions designed to facilitate and streamline manipulations of loop integrals and topologies. Even though the underlying algorithms are well-known to the practitioners and have already been implemented in many publicly available software packages, having them all conveniently accessible via high-level functions within one framework significantly lowers the bar for using those techniques in daily research. Owing to \textsc{FeynCalc}'s focus on flexibility, modularity and ease-of-use, users can casually employ these new functions whenever it is convenient for them, without the need to abandon their existing codes. The only condition is to convert the integral families appearing in the calculation into the \texttt{FCTopology}-notation, which normally can be done using just a few simple replacement rules.

Despite of all this progress, we would again like to stress that doing multiloop calculations with \textsc{FeynCalc} alone is not the goal we are aiming for. Our vision is to have a \textsc{FORM}-based calculational framework, where only certain steps (\eg topology minimization) should be performed using \textsc{FeynCalc}. To this end it was necessary to equip \textsc{FeynCalc} with the functions and symbols described in the present work. The next steps are to release an improved interface (a new version of \textsc{FeynHelpers}) connecting \textsc{FeynCalc} to other popular tools used in multiloop calculations and to make the related \textsc{FORM}-based setup publicly available. These tasks are currently being worked on and we hope to complete them in the near future.

\section*{Acknowledgments}

One of the authors (VS) would like to acknowledge Guido Bell, Simone Biondini, William Torres Boba\-dilla, David Broadhurst, Konstantin Chetyrkin, Joshua Davies, Florian Herren, Marvin Gerlach, Dennis Horstmann, Tobias Huber, Vitaly Magerya, Martin Lang, Fabian Lange, Ulrich Nierste, Erik Panzer, Kai Schönwald, Alexander Smirnov, Vladimir Smirnov and Matthias Steinhauser for useful discussions on different aspects of automatic perturbative calculations. The research of VS was supported by the Deutsche Forschungsgemeinschaft (DFG, German Research Foundation) under grant 396021762 — TRR 257 “Particle Physics Phenomenology after the Higgs Discovery”. This paper has been assigned preprint numbers TTP23-056, P3H-23-089 and SI-HEP-2023-27.

\appendix
\section{Pak's algorithm}

In this section we provide a brief description of Pak's algorithm including an illustrative example, similar to the one that was presented in ref.~\cite{Hoff:2016pot}. The starting point is always the derivation of the Symanzik polynomials $\mathcal{U}$ and $\mathcal{F}$ for the given integral or topology. Then we can construct a characteristic polynomial $\mathcal{P} \equiv \mathcal{U} + \mathcal{F}$\footnote{The choice 
$\mathcal{U} \times \mathcal{F}$ is also possible, but will usually contain a larger number of terms, so for performance reasons we prefer the sum and not the product.} describing the given integral family. In the case of a loop integral we also need to save the power of each denominator.

The polynomial is $\mathcal{P}$ typically of the form
\begin{equation}
	\mathcal{P} = \sum_{i=1}^l h_i \, x_1^{i_1} \cdot \ldots \cdot  x_n^{i_n},
\end{equation}
where $h_i$ is a kinematics-dependent factor, $n$ stands for the number of propagators or Feynman parameters $x_i$ and $l$ is the total number of terms.
It is convenient to write $\mathcal{P}$ as an  $l \times (n+1)$ matrix,
\begin{equation}
	\mathcal{P} \to \begin{pmatrix}
		h_1 & x_1^{a_1} & x_2^{a_2} & \cdots & x_n^{a_n} \\
		h_2 & x_1^{b_1} & x_2^{b_2} & \cdots & x_n^{b_n} \\
		\vdots & \vdots & \vdots & \cdots & \vdots \\
		h_l & x_1^{z_1} & x_2^{z_2} & \cdots & x_n^{z_n} \\
	\end{pmatrix},
\end{equation}
where we now want to find a canonical way to rename the $x_i$. To that aim we
set $i=1$ and generate new matrices by switching the $(i+1)$-th column with each of the next 
columns. We can keep track of these permutations by giving the matrices suitable
names containing the column numbers.

Looking only at the first $(i+1)$-columns of the new matrices we need to 
sort their rows. The exact nature of the sorting algorithm is irrelevant here, as long
as we always use the same procedure for all matrices. Computer algebra systems such
as \textsc{Mathematica} or  \textsc{Maple}  usually can sort lists out-of-the box. For
other programming languages one could \eg implement some lexicographic sorting algorithm.

Having obtained the sorted matrices we extract the $i$-th column from each of them, generating
a list of vectors. Upon sorting this list we take the first vector and keep
only matrices that contain the corresponding column while discarding the rest. Then we increase $i$ by one unit and start another iteration of the cycle, where we start
with the set of matrices obtained previously.

This procedure is repeated until we reach $i=n-1$. Then we collect the final permutations $\sigma$ of the remaining matrices, sort them and take the first permutation as our canonical way to name the Feynman parameters
$x_i$. Notice that $\sigma$ also provides us with a list of symmetries under $x_i$-renamings.

To illustrate this procedure let us consider the following characteristic polynomial
{\footnotesize
\begin{align}				
	\mathcal{P} = {\color{red} c_2 x_2 x_3}  + {\color{Turquoise}c_1 x_2^2} + {\color{blue} c_2 x_1 x_3} 
	+ {\color{brown} c_1 x_1^2} 
	\Rightarrow \begin{pmatrix} \color{red} c_2 & \color{red} 0 &  \color{red} 1 & \color{red} 1 \\ \color{Turquoise} c_1 & \color{Turquoise} 0 & \color{Turquoise} 2 & \color{Turquoise} 0 \\ \color{blue} c_2 & \color{blue} 1& \color{blue} 0 & \color{blue} 1 \\ 
		\color{brown} c_1 & \color{brown} 2& \color{brown} 0 & \color{brown} 0
	\end{pmatrix} \equiv M^{(123)}_0
	\vspace{-0.1cm}
\end{align}
}
In the first iteration ($i=1$) we start with $\{M^{(123)}_0\}$ and permute the second column
{\footnotesize
	\begin{align}
	\{M_0^{({\color{teal}1}{\color{orange} 2 }{\color{purple} 3})} &= 
	\begin{pmatrix}  
		c_2 & \color{teal} 0 & \color{orange} 1 & \color{purple} 1 \\ 
		c_1 & \color{teal} 0 & \color{orange} 2 & \color{purple} 0 \\ 
		c_2 & \color{teal} 1 & \color{orange} 0 & \color{purple} 1 \\
		c_1 & \color{teal} 2 & \color{orange} 0 & \color{purple} 0
	\end{pmatrix},
	M_0^{({\color{orange} 2 }{\color{teal}1}{\color{purple} 3})} = 
	\begin{pmatrix}  
		c_2 & \color{orange} 1 & \color{teal} 0 & \color{purple} 1 \\ 
		c_1 & \color{orange} 2 & \color{teal} 0 & \color{purple} 0 \\ 
		c_2 & \color{orange} 0 & \color{teal} 1 & \color{purple} 1 \\
		c_1 & \color{orange} 0 & \color{teal} 2 & \color{purple} 0
	\end{pmatrix},
	M_0^{({\color{purple} 3}{\color{orange} 2 }{\color{teal}1})} & = 
	\begin{pmatrix}  
		c_2  & \color{purple} 1  & \color{orange} 1 & \color{teal} 0 \\ 
		c_1  & \color{purple} 0  & \color{orange} 2 & \color{teal} 0 \\ 
		c_2   & \color{purple} 1 & \color{orange} 0 & \color{teal} 1 \\
		c_1 & \color{purple} 0  & \color{orange} 0 & \color{teal} 2
	\end{pmatrix}
	\}.
\end{align}
}
After sorting rows with respect to the first two columns we get
{\footnotesize
\begin{align}
	\{\tilde{M}_0^{({\color{teal}1}{\color{orange} 2 }{\color{purple} 3})} &= 
	\begin{pmatrix}  
		\bm c_1 &  \color{teal} \bm 0 & \color{orange} 2 & \color{purple} 0 \\ 
		\bm c_1 &  \color{teal} \bm 2 & \color{orange} 0 & \color{purple} 0 \\
		\bm c_2 &  \color{teal} \bm 0 & \color{orange} 1 & \color{purple} 1 \\ 
		\bm c_2 &  \color{teal} \bm 1 & \color{orange} 0 & \color{purple} 1
	\end{pmatrix},
	\tilde{M}_0^{({\color{orange} 2 }{\color{teal}1}{\color{purple} 3})} = 
	\begin{pmatrix}  
		\bm c_1 & \color{orange} \bm 0 & \color{teal} 2 & \color{purple} 0 \\
		\bm c_1 & \color{orange} \bm 2 & \color{teal} 0 & \color{purple} 0 \\ 
		\bm c_2 & \color{orange} \bm 0 & \color{teal} 1 & \color{purple} 1 \\
		\bm c_2 & \color{orange} \bm 1 & \color{teal} 0 & \color{purple} 1 
	\end{pmatrix}, 
	\tilde{M}_0^{({\color{purple} 3}{\color{orange} 2 }{\color{teal}1})} & = 
	\begin{pmatrix}  
		\bm c_1 & \color{purple} \bm 0  & \color{orange} 2 & \color{teal} 0 \\ 	
		\bm c_1 & \color{purple} \bm 0  & \color{orange} 0 & \color{teal} 2 \\
		\bm c_2 & \color{purple} \bm 1  & \color{orange} 1 & \color{teal} 0 \\ 
		\bm c_2 & \color{purple} \bm 1 & \color{orange} 0 & \color{teal} 1
	\end{pmatrix}
	\}.
\end{align}
}
The maximal vector among the second columns of all matrices is $({ 0,2,0,1})^T$,
which means that we need to keep $\tilde{M}_0^{({\color{teal}1}{\color{orange} 2 }{\color{purple} 3})}$ and 
$\tilde{M}_0^{({\color{orange} 2 }{\color{teal}1}{\color{purple} 3})}$ while discarding
$\tilde{M}_0^{({\color{purple} 3}{\color{orange} 2 }{\color{teal}1})}$.

The next iteration ($i=2$) starts with
{\footnotesize
\begin{align}
	\{
	\tilde{M}_0^{({\color{teal}1}{\color{orange} 2 }{\color{purple} 3})} &= 
	\begin{pmatrix}  
		c_1 &  \color{teal}  0 & \color{orange} 2 & \color{purple} 0 \\ 
		c_1 &  \color{teal}  2 & \color{orange} 0 & \color{purple} 0 \\
		c_2 &  \color{teal}  0 & \color{orange} 1 & \color{purple} 1 \\ 
		c_2 &  \color{teal}  1 & \color{orange} 0 & \color{purple} 1
	\end{pmatrix},
	\tilde{M}_0^{({\color{orange} 2 }{\color{teal}1}{\color{purple} 3})} = 
	\begin{pmatrix}  
		c_1 & \color{orange}  0 & \color{teal} 2 & \color{purple} 0 \\
		c_1 & \color{orange}  2 & \color{teal} 0 & \color{purple} 0 \\ 
		c_2 & \color{orange}  0 & \color{teal} 1 & \color{purple} 1 \\
		c_2 & \color{orange}  1 & \color{teal} 0 & \color{purple} 1 
	\end{pmatrix}
	\}.
\end{align}
}
Permuting the third column we get
{\footnotesize
\begin{align}
		\{
		{M}_1^{({\color{teal}1}{\color{orange} 2 }{\color{purple} 3})} &= 
		\begin{pmatrix}  
			c_1 &  \color{teal}  0 & \color{orange} 2 & \color{purple} 0 \\ 
			c_1 &  \color{teal}  2 & \color{orange} 0 & \color{purple} 0 \\
			c_2 &  \color{teal}  0 & \color{orange} 1 & \color{purple} 1 \\ 
			c_2 &  \color{teal}  1 & \color{orange} 0 & \color{purple} 1
		\end{pmatrix},
		{M}_1^{({\color{teal}1}{\color{purple} 3}{\color{orange} 2 })} = 
		\begin{pmatrix}  
			c_1 &  \color{teal}  0 & \color{purple} 0 & \color{orange} 2  \\ 
			c_1 &  \color{teal}  2 & \color{purple} 0 & \color{orange} 0  \\
			c_2 &  \color{teal}  0 & \color{purple} 1 & \color{orange} 1  \\ 
			c_2 &  \color{teal}  1 & \color{purple} 1 & \color{orange} 0 
		\end{pmatrix}, 
		{M}_1^{({\color{orange} 2 }{\color{teal}1}{\color{purple} 3})} = 
		\begin{pmatrix}  
			c_1 & \color{orange}  0 & \color{teal} 2 & \color{purple} 0 \\
			c_1 & \color{orange}  2 & \color{teal} 0 & \color{purple} 0 \\ 
			c_2 & \color{orange}  0 & \color{teal} 1 & \color{purple} 1 \\
			c_2 & \color{orange}  1 & \color{teal} 0 & \color{purple} 1 
		\end{pmatrix}, \nonumber \\ 
		{M}_1^{({\color{orange} 2 }{\color{purple} 3}{\color{teal}1})}  &= 
		\begin{pmatrix}  
			c_1 & \color{orange}  0 & \color{purple} 0 & \color{teal} 2  \\
			c_1 & \color{orange}  2 & \color{purple} 0 & \color{teal} 0  \\ 
			c_2 & \color{orange}  0 & \color{purple} 1 & \color{teal} 1  \\
			c_2 & \color{orange}  1 & \color{purple} 1 & \color{teal} 0  
		\end{pmatrix}
		\}.
\end{align}
}
Sorting rows with respect to the first three columns does not introduce any changes in the matrices

\begin{align}
	\{
	\tilde{M}_1^{({\color{teal}1}{\color{orange} 2 }{\color{purple} 3})} &= 
	{M}_1^{({\color{teal}1}{\color{orange} 2 }{\color{purple} 3})},
	\tilde{M}_1^{({\color{teal}1}{\color{purple} 3}{\color{orange} 2 })} = 
	{M}_1^{({\color{teal}1}{\color{purple} 3}{\color{orange} 2 })},
	\tilde{M}_1^{({\color{orange} 2 }{\color{teal}1}{\color{purple} 3})}  = 
	{M}_1^{({\color{orange} 2 }{\color{teal}1}{\color{purple} 3})}, 
	\tilde{M}_1^{({\color{orange} 2 }{\color{purple} 3}{\color{teal}1})}  = 
	{M}_1^{({\color{orange} 2 }{\color{purple} 3}{\color{teal}1})}
	\}
\end{align}
This time the maximal vector the third columns is $({ 2,0,1,0})^T$ so that
we  keep only $\tilde{M}_1^{({\color{teal}1}{\color{orange} 2 }{\color{purple} 3})}$ and 
$\tilde{M}_1^{({\color{orange} 2 }{\color{teal}1}{\color{purple} 3})}$. Since $i=3 = n-1 = 3$ the algorithm terminates here.

The outcome of this procedure are the symmetries under the renamings of $x_i$ 
\begin{equation}
	\sigma = \{(123), (213)\},
\end{equation}
meaning that
\begin{align}
	\mathcal{P}^{(123)} &= c_2 x_2 x_3  +c_1 x_2^2 + c_2 x_1 x_3 + c_1 x_1^2, \\
	\mathcal{P}^{(213)} &= c_2 x_1 x_3  +c_1 x_1^2 + c_2 x_2 x_3 + c_1 x_2^2
\end{align}
are equivalent. Here $(123)$ is our canonical naming scheme. With $\mathcal{P}^{(123)}$
we have an expression that uniquely characterizes the corresponding set of propagators. Any other
loop integral or integral topology that differs from the given one only by a finite set of loop momentum shifts
should have the same characteristic polynomial. This is why by  comparing $\mathcal{P}$'s (and propagator
powers) we can identify one-to-one mappings between different integrals.

\section{Mapping of smaller topologies into larger topologies}

Pak algorithm can only find mappings between topologies that contain the same number of propagators. 
In practice, one often encounters cases were a topology with a smaller number of propagators happens to fit into
a topology with a larger number of propagators. Such relations can be uncovered in the following way.

First, we need a list of parent topologies that contain enough propagators to allow fitting
smaller topologies. Those parent topologies can have external origin or stem from the current calculation.
Ideally, each parent topology should have a complete set of propagators forming a basis, so that it can
be directly used for IBP reduction. In the next step, we can analyze each parent topology and determine
all of its nonvanishing subtopologies. This means that we try removing one, two or more propagators from that
topology and then check if the resulting topology becomes scaleless and hence vanishes\footnote{The criterion
for checking the scalefulness of an arbitrary loop integral was presented in ref.~\cite{Pak:2010pt}. A
good description of the algorithm can be found in Sec.~2.3.4 of ref.~\cite{Hoff:2015kub}. In \textsc{FeynCalc}
the corresponding routines are called \texttt{FCLoopScalelessQ} (for checking loop integrals or topologies) and
\texttt{FCLoopPakScalelessQ} (for checking characteristic polynomials $\mathcal{P}$).}. This can be done
using the routine \texttt{FCLoopFindSubtopologies}. Each nonvanishing subtopology receives a marker of the form
\texttt{FCGV[\-"Subtopology\-Of"] -> topologyID} which is added to the 6th slot of the corresponding \texttt{FCTopology} object,
where \texttt{topologyID} refers to the parent topology from which this subtopology has been obtained.

Then, we can run \texttt{FCLoopFindTopologyMappings} where the smaller topologies are given as the first
argument of the function, while the list of subtopologies obtained from \texttt{FCLoop\-Find\-Subtopologies}
is passed via the option \texttt{Pre\-ferred\-Topologies}. Mappings between small topologies and nonvanishing
subtopologies with the same number of propagators can be now determined using the conventional Pak algorithm.
Once we have the momentum shifts that allow us to map a small topology into a nonvanishing subtopology of a 
larger topology, we can immediately work out the mapping into the parent topology. Thanks to the 
\texttt{SubtopologyOf}-marker we already know the name of the parent topology, so we only need to apply
the momentum shifts to it and pass both topologies to the 
auxiliary routine \texttt{FCLoopCreateRuleGLIToGLI}. This gives us the desired relation between a smaller
and a larger topology.

One should remark, that the determination of nonvanishing subtopologies can be quite time-consuming and 
tends to generate large numbers of resulting topologies. Thus, given a big set of complicated topologies, the total number of their nonvanishing subtopologies can easily get several orders of magnitude larger. Processing all those topologies using \texttt{FCLoopFindTopologyMappings} can, therefore, take a lot of time. For this reason we do not recommend using this feature extensively, unless the number of topologies is small (\eg  $\mathcal{O}(100)$) or it is absolutely necessary to find some specific mappings. However, we aim to speed up this functionality in near future.

\section{Limitations of the current approach to topology minimization} \label{sec:mixedprops}

While Pak algorithm can tell us that two topologies are identical, finding a set of momentum shifts that realizes
this mapping may not always be straightforward. 

For topologies containing only standard quadratic propagators of the form\footnote{Here and below it is understood that $m$ can be also zero without changing the presented discussion.} $l^2 \pm m^2$ the situation is very simple. The 
line momentum $l$, which is a linear combination of some loop momenta $p_j$ and external momenta $q_k$, directly gives us the 
momentum flow through the corresponding propagator. The Pak algorithm orders 
the Feynman parameters of each topology in a canonical way and this ordering can be applied also to the propagators since
each Feynman parameter $x_i$ corresponds to the $i$th propagator of the integral family. Having two sets of ordered propagators
$\{D_1, D_2, \ldots, D_n\}$ and $\{D'_1, D'_2, \ldots, D'_n\}$ with $D^{(')}_i = {l^{(')}}_i^2 \pm {m^{(')}}^2_i$ we can directly
write down a system of equations\footnote{We square both sides of the equation to allow for solutions that introduce sign changes of some loop momenta \eg $p_j \to - p_j$.} for the line momenta
\begin{equation}
l_1^2 = (l_1')^2, \quad l_2^2 = (l_2')^2, \quad  \ldots, \quad l_n^2 = (l_n')^2
\end{equation}
and solve it for $p_k$ or $p'_k$. Since both topologies are identical, there must be at least one solution, which gives us the 
desired momentum shifts. Notice, that in general such systems turn out to be overdetermined since the number of loop momenta
is usually much smaller than the total number of propagators.

Now let us consider topologies containing some eikonal propagators $E_i$. Pak algorithm still can recognize that they are identical,
as eikonal propagators do not pose any additional complications when calculating the Symanzik polynomials $\mathcal{U}$ and $\mathcal{F}$.
However, denominators of the form $q \cdot s \pm m^2$ do not allow us to recover the momentum flow through this propagator unambiguously. This complicates
the process of determining the necessary momentum shifts. To avoid dealing with other than linear systems of equations, we choose a pragmatic approach, where we simply remove the eikonal propagators from both sets of ordered propagators.

Since every loop integral containing only eikonal propagators will be scaleless, there must be at least one quadratic propagator $D_i$
for each loop momentum. Furthermore, as have been observed above, the system of equations we need to solve is usually overdetermined so
that upon removing suitable equations it should still remain solvable. Although we recognize that there may be pathological cases where
this approach will fail, as of now we are not aware of a better solution to this problem that does not involve massive performance penalties.

Last but not least, there also exists an interpolating case between quadratic and purely eikonal propagators, where the denominators are of the form
$l^2 + c_1 l \cdot s + c_2$ with $c_i$ being some constants. Such propagators can arise \eg when doing asymptotic expansions at the integrand
level. On the one hand, naively discarding them may lead to unsolvable systems of equations. On the other hand, automatic determinations of the momentum
flow through such lines can be tricky. To this aim \textsc{FeynCalc} contains a helper function 
\texttt{FC\-Loop\-Replace\-Quadratic\-Eikonal\-Propagators} that should be applied to a list of topologies containing mixed propagators. Given the loop momenta as well as some extra replacement rules\footnote{\eg that
$p_1^2 - 2 p_1 \cdot p_2 + p_2^2$ combines to $(p_1 - p_2)^2$} this routine should be able to rewrite mixed propagators in terms of purely quadratic ones.

\bibliographystyle{elsarticle-num}
\bibliography{final.bib}

\end{document}